\documentclass[10pt,conference]{IEEEtran}
\IEEEoverridecommandlockouts
\usepackage{cite}
\usepackage{amsmath,amssymb,amsfonts}
\usepackage{algorithmic}
\usepackage{graphicx}
\usepackage{textcomp}
\usepackage{xcolor}
\usepackage{booktabs}
\usepackage{subcaption}
\usepackage{braket}
\usepackage{tikz}
\usetikzlibrary{svg.path}
\usepackage{scalerel}

\usepackage[hyperfootnotes=false,hidelinks]{hyperref}

\def\BibTeX{{\rm B\kern-.05em{\sc i\kern-.025em b}\kern-.08em
    T\kern-.1667em\lower.7ex\hbox{E}\kern-.125emX}}
\begin{document}

\newcommand{\arxiv}[1]{#1}
\newcommand{\acronym}{INJEQT}
\newcommand{\tdg}{TDG}
\newcommand{\rz}{Rz}
\newcommand{\idle}{\mathcal{I}}
\newcommand{\inmodule}{\mathcal{B}}
\newcommand{\intermodule}{\mathcal{C}}
\newcommand{\shift}{\mathcal{U}}
\newcommand{\tprep}{\mathcal{F}}
\newcommand{\tstate}{\ket{T}}
\newcommand{\pauliphi}{P(\varphi)}
\makeatletter
\newcommand{\linebreakand}{%
  \end{@IEEEauthorhalign}
  \hfill\mbox{}\par
  \mbox{}\hfill\begin{@IEEEauthorhalign}
}
\makeatother

\definecolor{orcidlogocol}{HTML}{A6CE39}
\tikzset{
  orcidlogo/.pic={
    \fill[orcidlogocol]
    svg{M256,128c0,70.7-57.3,128-128,128C57.3,256,0,198.7,0,128C0,57.3,57.3,0,128,0C198.7,0,256,57.3,256,128z};
    \fill[white] svg{M86.3,186.2H70.9V79.1h15.4v48.4V186.2z}
    svg{M108.9,79.1h41.6c39.6,0,57,28.3,57,53.6c0,27.5-21.5,53.6-56.8,53.6h-41.8V79.1z
    M124.3,172.4h24.5c34.9,0,42.9-26.5,42.9-39.7c0-21.5-13.7-39.7-43.7-39.7h-23.7V172.4z}
    svg{M88.7,56.8c0,5.5-4.5,10.1-10.1,10.1c-5.6,0-10.1-4.6-10.1-10.1c0-5.6,4.5-10.1,10.1-10.1C84.2,46.7,88.7,51.3,88.7,56.8z};
  }
}

\newcommand\orcidicon[1]{\textsuperscript{\href{https://orcid.org/#1}{\mbox{\scalerel*{
          \begin{tikzpicture}[yscale=-1,transform shape]
            \pic{orcidlogo};
          \end{tikzpicture}
}{|}}}}}

\title{\acronym{}: Improved Magic-State \underline{Inje}ction Protocol for Fault-Tolerant \underline{Q}uantum Ex\underline{t}ractor Architectures
}

\author{
\IEEEauthorblockN{
Sayam Sethi\IEEEauthorrefmark{3}\IEEEauthorrefmark{1}\orcidicon{0009-0005-3056-5285},
Sahil Khan\IEEEauthorrefmark{2}\orcidicon{0009-0000-4160-8010},
Aditi Awasthi\IEEEauthorrefmark{3}\orcidicon{0009-0006-7768-8503},
Abhinav Anand\IEEEauthorrefmark{2}\IEEEauthorrefmark{4}\orcidicon{0000-0002-8081-2310},
Jonathan Mark Baker\IEEEauthorrefmark{3}\orcidicon{0000-0002-0775-8274}
}
\IEEEauthorblockA{
\IEEEauthorrefmark{3}Electrical and Computer Engineering, The University of Texas at Austin\\
\IEEEauthorrefmark{2}Electrical and Computer Engineering, Duke University\\
\IEEEauthorrefmark{4}Lawrence Berkeley National Laboratory\\
\IEEEauthorrefmark{1}\href{mailto:sayams@utexas.edu}{sayams@utexas.edu}
}
}

\maketitle

\begin{abstract}
Near-term Fault-Tolerant Quantum Computing (FTQC) system designs are constrained by limited error budgets and the largely sequential execution of non-Clifford operations. As a result, reducing the number of the most-error prone instructions becomes critical for successful program execution. In this work, we study the extractor architecture~\cite{he_extractors_2025,yoder_tour_2025}, a recently proposed FTQC design that enables universal quantum computation on spatially-efficient quantum error correcting codes (QECC) such as the bivariate-bicycle code family~\cite{bravyi_high-threshold_2024}. In these architectures, over $90\%$ of the total program error arises from the synthillation~\cite{campbell_unifying_2017} process, which involves $\tstate$-state preparation and injection to implement non-Clifford operations. We observe that standard \rz{} synthillation requires multiple sequential $\tstate$-state injections, each incurring an inter-module measurement, the most expensive instruction in the architecture~\cite{yoder_tour_2025}, which cumulatively dominate the overall error budget. 

To address this bottleneck, we propose \acronym{}, a $2$-factory design that uses an auxiliary code capable of synthesizing $\rz(\theta)$ states with lower error rates. These states are then injected into the extractor modules using only a constant number of inter-module measurements. This approach reduces overall error rates by up to $22\times$. While a naive implementation of this design leads to an increase in execution time, we reduce the time overhead by a pre-fetching strategy that prepares the \rz{} states and their correction states in parallel. This approach improves the wall-clock time by up to $13\times$ and reduces the space-time cost by up to $7.2\times$, for an optimal choice of the number of \acronym{} factories for each metric. We evaluate the performance of \acronym{} for multiple state preparation techniques such as distillation, cultivation and STAR, and model the execution times for both lattice surgery-based injections and transversal CNOT based injections. Our results demonstrate that \acronym{} is robust across factory choices and device technologies, enabling more efficient architectural designs for FTQC.
\end{abstract}


\section{Introduction}

Quantum computers enable wide-ranging applications across a range of scientifically important problems, including factoring, optimization, and quantum simulations~\cite{shor1999polynomial, kivlichan2020qsimelectrons, childs2018firstqsim, harrow2009hhl, reiher2017nitrogenfixation}. The qubits used to perform computation are highly susceptible to errors; therefore, quantum error correction (QEC) must be implemented to ensure successful program execution. QEC codes use many noisy physical qubits to encode information into reliable \textit{logical} qubits, enabling fault-tolerant quantum computation (FTQC). However, this comes at an overhead in terms of both space (number of qubits) and time (program execution duration). Recent works have proposed the bivariate bicyclic code family~\cite{bravyi_high-threshold_2024}, which is a spatially-efficient code family of the larger qLDPC (quantum Low-Density Parity Check) code family.

Performing universal fault-tolerant computation on this code family requires a new architecture, known as the extractor-based architecture~\cite{swaroop_universal_2024,he_extractors_2025,yoder_tour_2025}, which requires additional components such as adapters, Logical Processing Unit (LPUs), and resource-state (or magic-state) factories. The resource-state factory produces specialized non-Clifford resource states such as $\tstate$, or $\rz(\varphi)\ket{+}$ for $\varphi \in [0,2\pi)$. This enables universal quantum computation by enabling $\pauliphi$ rotations in the Pauli-Based computing model. Such resource states cannot be generated transversally on the base extractor code, and must be prepared fault-tolerantly through protocols such as \textit{distillation} \cite{litinski_magic_2019}, \textit{cultivation} \cite{gidney_magic_2024,sahay_fold-transversal_2026}, or direct \rz{} synthesis~\cite{akahoshi_partially_2023,toshio_practical_2024,toshio_star-magic_2026,yoshioka_transversal_2025} in a different code, which is then injected onto the base code. These state preparation procedures are both resource-intensive and nondeterministic. Despite the choice of protocol used to generate the resource state, the \textit{injection} of this state into the qLPDC code is typically the most erroneous operation~\cite{sethi_optimizing_2026,yoder_tour_2025} since it requires the more expensive inter-module measurements. The resource-state preparation along with injection is known as synthillation~\cite{campbell_unifying_2017}. On average, we find that over $90\%$ of the total program infidelity can be attributed to this synthillation process.

\begin{figure*}
    \centering
    \includegraphics[width=\linewidth]{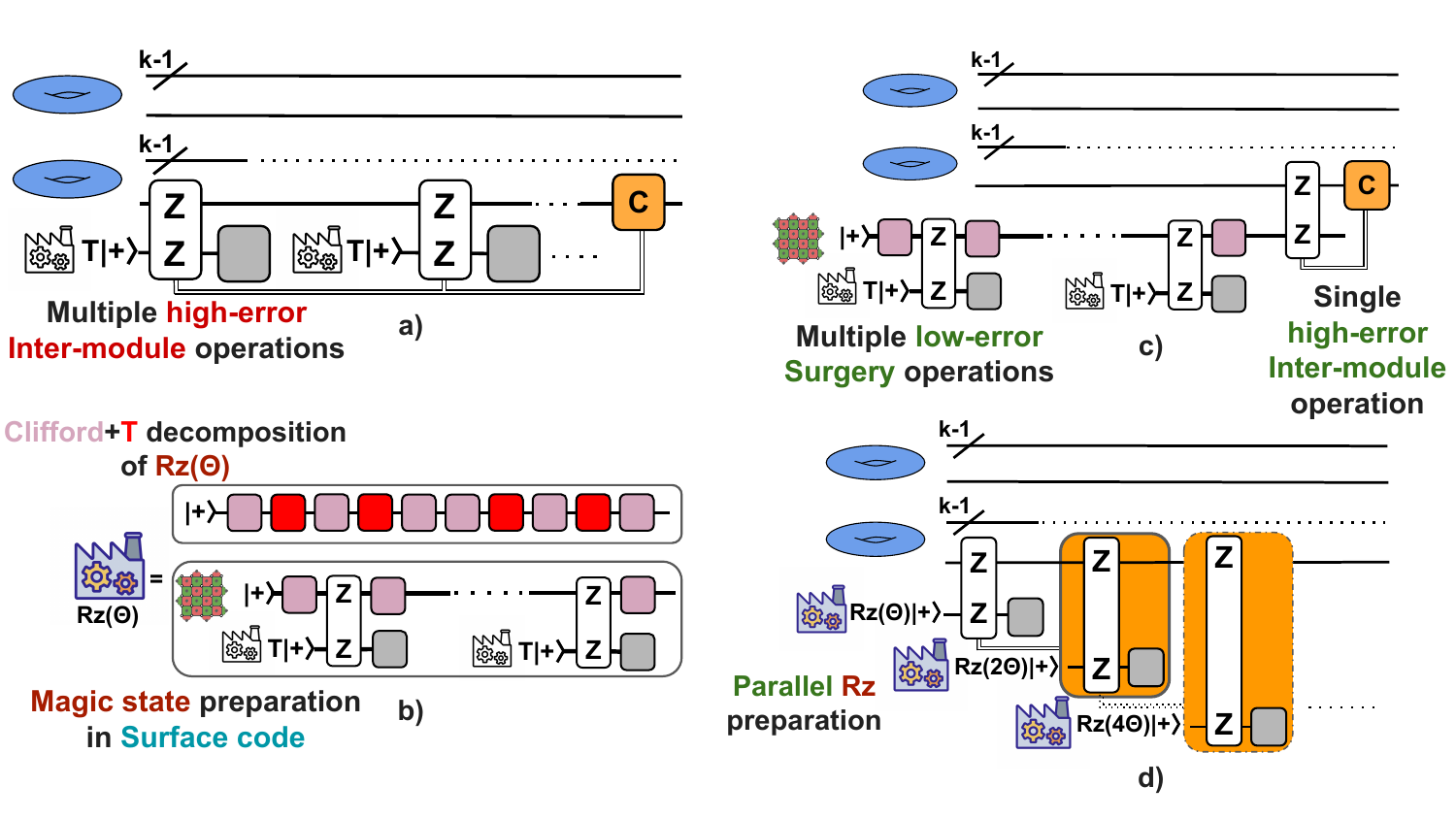}
    \caption{Overview of magic state injection techniques. (a) Baseline standard injection: Represented by the Tour de Gross architecture, which relies on multiple high-error inter-module operations. (b) Surface code \rz-factory: Illustration of the Clifford+T decomposition of the \rz-state and the synthesis of an \rz-factory within the surface code using multiple $\tstate$-factories. (c) \textbf{Proposed qLDPC injection}: Our method injects the prepared \rz-state into the qLDPC code, utilizing a single inter-module operation and fault-tolerant lattice surgery operations on surface code. (d) \textbf{Time-efficient injection (\acronym{})}: Simultaneous preparation of multiple \rz-states required for post-injection corrections, yielding a highly time-efficient execution.
}
    \label{fig:intro-fig}
\end{figure*}

Prior work has outlined how to perform magic state injection within extractor architectures, including techniques for parallel injections and logical qubit mappings~\cite{yoder_tour_2025,xu_distilling_2026,khan_architecting_2026,sethi_optimizing_2026}. However, they assume a static error cost and do not explore the space-time tradeoffs of synthillation. In this work, we focus on reducing the error overhead by introducing additional space-time tradeoffs to lower the overall application error.
We propose \textbf{\acronym{}}, an enhancement to traditional magic state injection using extractors by introducing another intermediary code which has a lower synthesis error for a given $\pauliphi$ rotation. To mitigate additional time overhead for this synthesis, we propose a two-level pipelined factory approach. We outline our approach in Figure~\ref{fig:intro-fig}, wherein we trade-off the more erroneous inter-module measurements for less erroneous auxiliary code operations. A naive implementation of this approach is time-inefficient. We address this issue by proposing a pre-fetching technique that prepares multiple of these resource states in parallel. We find that \acronym{} obtains up to a $13\times$ improvement on execution time (wall-clock time) while also reducing the program error by up to $22\times$. Our $2$-level factory adds only a $25\%$ increase in space in the worst case, with a modest increase on average ($3-10\%$, depending on the choice of auxiliary code and factory). We evaluate \acronym{} with multiple different factories and error rates, and also compare trade-offs with different injection protocols on the auxiliary code. We find that the surface code is a natural choice for the auxiliary code to balance space and time tradeoffs, especially in the context of synthillation. We also find that experimentally, the number of \acronym{} factories needed depends on the benchmark and the metric being optimized (such as wall-clock time or space-time cost). Our contributions are as follows:
\begin{enumerate}
    \item We find that intermodule measurements during synthillation contributes the majority of the error cost. We reduce this overhead by introducing a $2$-level factory which synthesizes the intermediate \rz{} resource state in the auxiliary code. This approach improves the program error rates by about $11\%$ on average for distillation, and by about $11\times$ for cultivation.
    \item We further improve the time overheads of our $2$-level factory by proposing a pre-fetching approach that prepares multiple \rz{} states (including corrections) in parallel, and in advance. This approach reduces the wall-clock time of program execution by $7\times$ on average, with a $2\times$ average improvement in space-time.
    \item We rigorously study the sensitivity of \acronym{} to multiple factory protocols (distillation, cultivation, STAR), and also evaluate different auxiliary code execution models, showing that \acronym{}'s techniques are robust to factory and device technology.
\end{enumerate}


\section{Background}
\subsection{Quantum Error Correcting Codes}
Quantum error correcting codes abstract multiple physical qubits into logical qubits. A quantum error correcting code is parametrized by $[[n,k,d]]$, where $n$ is the number of data qubits which collectively hold program information, $k$ is the number of logical qubits which are the total number of program qubits, and $d$ is the code distance which affects the amount of correctable errors. Quantum error correcting codes interleave syndrome extraction (which protects idle quantum memory) with fault tolerant logical operations. Logical operations can be done in a \textit{transversal} manner, where a logical operation is achieved through broadcasting the same operation among physical qubits, or using code surgery where multi-body projective measurements affect the logical state of the code. In this paper, we choose one representative code from each logical operation paradigm, the surface code~\cite{litinski_game_2018} and the gross code~\cite{bravyi_high-threshold_2024,yoder_tour_2025}, representing the transversal and code surgery based approaches, respectively. The transversal logical operation set for the surface code include H, S, and CX~\cite{litinski_lattice_2018}. The gross code admits a separate fault tolerant instruction set detailed in Section~\ref{sec:extractor-isa}.

\subsubsection{Surface Code}
The surface code with code parameters $[[d^2, 1, d]]$ is among the most extensively studied quantum error correcting codes due to its simple planar topology and high \textit{threshold} (i.e., error tolerance). Syndrome extraction (logical idling operation) in the surface code requires $O(d)$ cycles, where each cycle consists of measuring all Pauli stabilizers in the code. 

\subsubsection{Gross Code}
A well-studied high-rate code that we use throughout this paper is an instance of the Bivariate Bicycle family with parameters $[[144,12,12]]$, referred to as the gross code~\cite{bravyi_high-threshold_2024}. The gross code uses a defined compute model known as the extractor based model~\cite{he_extractors_2025}. Explicit instructions for performing logical computation are laid out in the Tour de Gross (\tdg{}) architecture~\cite{yoder_tour_2025}. To the best of our knowledge, the well defined instructions laid out in \cite{yoder_tour_2025} is the only surgery-based compute model for high-rate qLDPC codes. We use this as a reference for instruction error rates and timings. Further details are provided in Section~\ref{sec:extractor-isa}.

\subsection{Pauli-Based Computing}\label{sec:pbc}

Pauli-Based Computing (PBC) is a model of quantum computation in which operations are restricted to measurements of multi-qubit Pauli operators, together with classical feed-forward operations conditioned on measurement outcomes. Instead of applying arbitrary unitary gates, computation proceeds by adaptively selecting Pauli measurements whose outcomes both drive the evolution of the quantum state and determine subsequent measurement choices. This measurement-driven paradigm is closely related to measurement-based quantum computing, but is distinguished by its restriction to the Pauli group, which enables efficient classical tracking of quantum states via stabilizer formalisms.

\subsubsection{Eastin--Knill Theorem}\label{sec:eastin-knill}
The Eastin--Knill theorem~\cite{eastin_restrictions_2009} establishes a fundamental limitation on the structure of fault-tolerant quantum computation. It states that no quantum error-correcting code can support a universal set of logical gates implemented entirely by transversal operations. As a consequence, at least one non-transversal operation must be introduced to achieve universality. In practice, this limitation is overcome by techniques such as magic state distillation and injection, code switching, or lattice surgery, which enable the fault-tolerant implementation of a universal set of logical operations.
The relevance of the Eastin--Knill theorem to Pauli-Based Computing lies in its implication that Pauli and Clifford operations alone are insufficient for universal computation. To achieve universal quantum computation, PBC must incorporate non-Clifford resources, typically in the form of specially prepared resource states (e.g., magic states). These states are prepared using magic state preparation techniques, including \textit{cultivation}~\cite{gidney_magic_2024}, \textit{distillation}~\cite{litinski_magic_2019}, and the \textit{STAR} architecture~\cite{akahoshi_partially_2023}. Once prepared, magic states must be injected into the corresponding logical program qubits to enable universal fault tolerant computation. A standard fault tolerant workflow thus incorporates \textit{magic state injection} along with logical Clifford operations while magic state production proceeds in parallel.

\subsection{Magic-State Injection}\label{sec:ms-injection}

Magic state injection can be performed in two ways: \textit{surgery based injection} (Figure \ref{fig:lattice-surgery-injection}) or \textit{transversal based injection} (Figure \ref{fig:transversal-injection}). The chosen logical compute paradigm (transversal, surgery based) determines the method of injection and slightly affects both the error rates and associated time cost of the injection operation. Injection via lattice surgery takes $d \times s$ operations~\cite{litinski_game_2018}, whereas injection via transversal operations takes $s + 1$~\cite{cain_correlated_2024}, where $d$ is the code distance and $s$ is the syndrome extraction time. For the surface code, we have $s = 6$. The transversal method requires no additional qubit overhead, whereas the lattice surgery method requires an additional ancillary qubits. The cost scales on the order of $O(d^2)$~\cite{kaavyatcnot}.
For hardware with fixed connectivity constraints, transversal operations are often impractical due to the need for long range interactions. Consequently, superconducting devices typically rely on surgery based compute model which is less demanding on connectivity. In contrast, neutral atom and trapped ion platforms, with less restrictive connectivity, can employ \textit{either} of the surgery or transversal compute model. However, shuttling the qubits around to meet the connectivity demand can still incur a non-negligible cost.

\begin{figure}
    \centering
    \begin{subfigure}{\linewidth}
        \centering
        \includegraphics[width=0.7\linewidth]{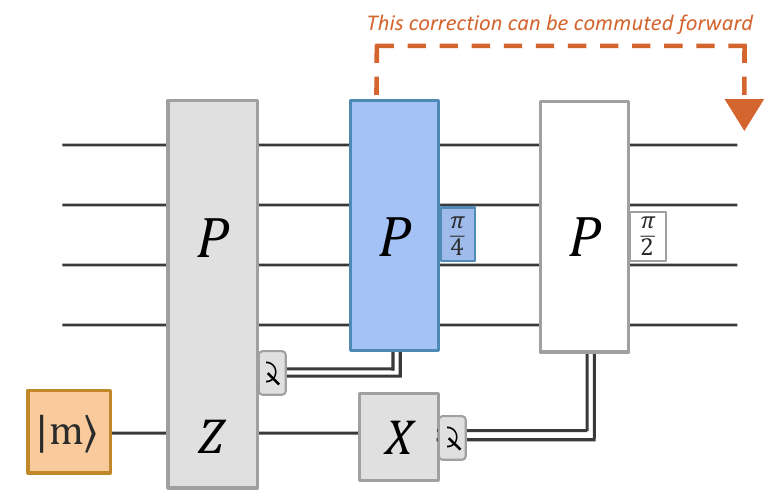}
        \caption{Lattice surgery injection (the Clifford $\pi/4$ correction can be commuted forward and is not executed).}\label{fig:lattice-surgery-injection}
    \end{subfigure}
    \begin{subfigure}{\linewidth}
        \centering
        \includegraphics[width=0.8\linewidth]{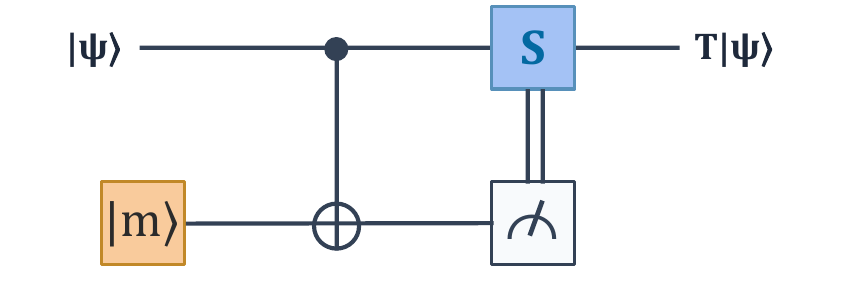}
        \caption{Transversal CNOT injection}\label{fig:transversal-injection}
    \end{subfigure}
    \caption{The two magic-state injection techniques considered in this work. $\ket{m}$ is the magic state being injected which can be either of $T\ket{+}$ or $\rz(\theta)\ket{+}$ depending on the factory choice.}\label{fig:ms-injections}
\end{figure}

\section{Extractor-Based Architectures}

\subsection{Extractor Instruction Set Architecture}\label{sec:extractor-isa}
\begin{figure}
    \centering
    \includegraphics[width=0.9\linewidth]{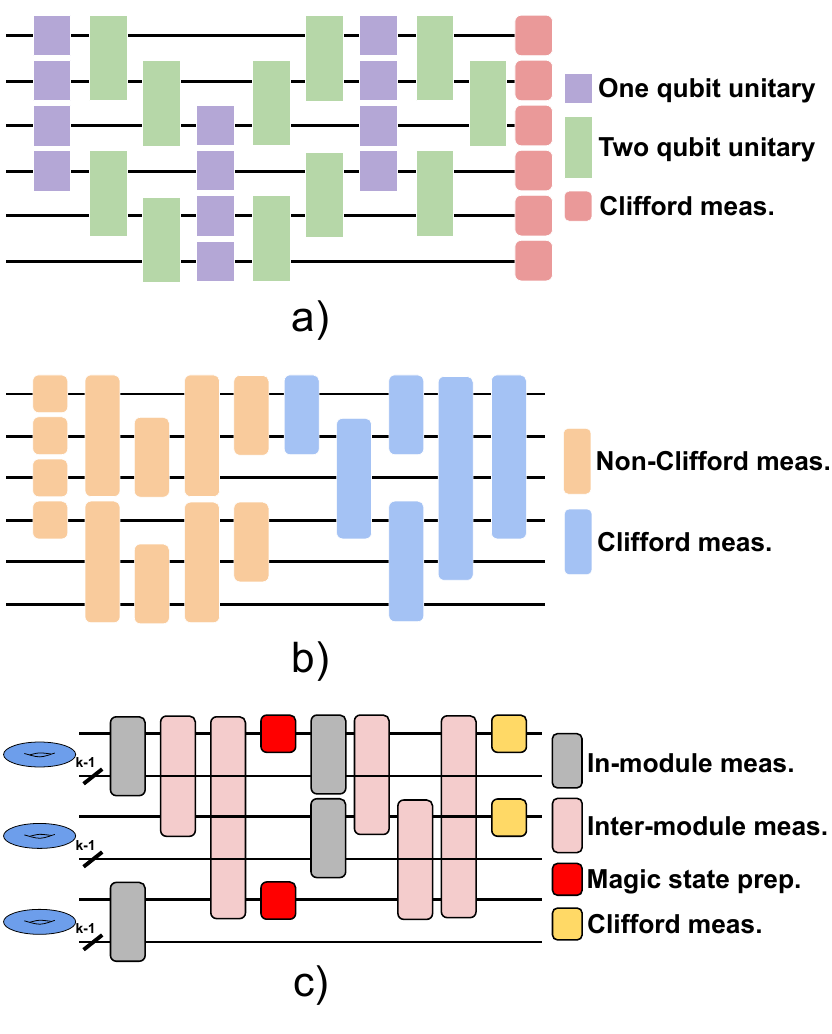}
    \caption{Illustration of circuits across different abstraction layers. (a) Standard Circuit Model illustrating single and two-qubit unitary gates followed by measurements. (b) Pauli-Based Computing Model representing the computation as a sequence of Clifford and non-Clifford measurements. (c) Tour de Gross ISA, showing the modular instruction set architecture including In-module, Inter-module measurement and magic state preparation. For further details on the exact compilation process, see Ref.~\cite{litinski_game_2018} and Ref.~\cite{yoder_tour_2025}.}
    \label{fig:extractor-compilation}
\end{figure}

Compiling to the Extractor Instruction Set Architecture (ISA) first involves compiling an input Clifford + \rz{} circuit to a Pauli-Based circuit in terms of $\pauliphi$ gates followed by measurements (Section~\ref{sec:pbc}). We then compile these Pauli-based gates into their corresponding ISA instructions: idle ($\idle$), in-module measurements ($\inmodule$), inter-module measurements ($\intermodule$), shift automorphisms ($\shift$), and magic-state preparation ($\tprep$). We describe the compilation pipeline in Figure~\ref{fig:extractor-compilation}. We also detail the error rates and execution times for the gross code in Table~\ref{table:tdg-stats} using data from Ref.~\cite{yoder_tour_2025}.

\begin{table}[htbp!]
    \setlength\tabcolsep{0pt}
    \centering
    \begin{tabular*}{\columnwidth}{@{\extracolsep{\fill}}llrr}
        \toprule
        Instruction & Notation & Steps ($\tau_\mathsf{op}$) & Logical error rate ($\varepsilon_\mathsf{op}$) \\
        \midrule
        idle & $\idle$& $8$ & $10^{-14.8\pm 0.4}$\\
        shift automorphism & $\shift$ & $14$ & $10^{-12.2\pm 0.5}$\\
        in-module measurement & $\inmodule$ & $120$ & $10^{-9.0\pm 0.2}$\\
        inter-module measurement & $\intermodule$ & $120$ & $10^{-7.4\pm 0.3}$\\
        \bottomrule
    \end{tabular*}
    \caption{Logical error rates and time steps at physical error rate $p = 10^{-4}$ for \tdg{}~\cite{yoder_tour_2025}.}\label{table:tdg-stats}
\end{table}

\subsection{Magic-State Injection in the Extractor Architecture}

All ISA instructions detailed in Section~\ref{sec:extractor-isa}, except for magic-state preparation, are native to the qLDPC code (which we refer to as code module) used by the extractor architecture (e.g. the gross code). Magic-state preparation, however, requires a separate magic-state factory. The factory prepares the state in a different code and the prepared state is injected into the module via an inter-module measurement between the factory and the module. Therefore, the choice of factory and injection protocol determines the execution time, fidelity, and the space cost of the quantum application being executed on the extractor architecture.

\par We report the different kinds of factories considered in this work along with their associated parameters in Table~\ref{table:factory-error-rates}. Note that the code distances were selected to ensure that the factory LERs are comparable to or better than the inter-module measurement logical error rates reported in Table~\ref{table:tdg-stats}. This is because 1. each preparation is coupled with an injection, thereby making the total error rate of factory injects be dominated by the more expensive of the two, which necessitates that factories have comparable LERs, and/or 2. inter-module measurements operate between different codes, which are typically placed on distinct quantum processes and connected via photonic/optical inter-connects which makes them more erroneous~\cite{jacinto_network_2026,liu_remote_2026,sinclair_fault-tolerant_2025}. The error rate for cultivation is much lower than the error rates for other factories, since $d=3$ colour code does not suffice to achieve comparable error rates (they saturate at $\sim10^{-6}$) to $\varepsilon_\intermodule$ at $p = 10^{-4}$~\cite{gidney_magic_2024,sahay_fold-transversal_2026}, however, using $d = 5$ colour code ends up being \textit{too good}.

\begin{table}
    \setlength\tabcolsep{0pt}
    \centering
    \begin{tabular*}{\columnwidth}{@{\extracolsep{\fill}}llrrrr}
        \toprule
         Factory & Type & Code Distance & LER ($\varepsilon_\tprep$) & $\mathbb{E}\left[\mathrm{Steps}\right]$ & $\#q$\\
         \midrule
         Distillation~\cite{yoder_tour_2025} & $\tstate$ & $d_X = 7, d_Z = d_M = 3$ & $4.4\times 10^{-8}$ & $108.6$ & 810\\
         Cultivation~\cite{gidney_magic_2024} & $\tstate$ & $d_\mathrm{CC} = 5, d_\mathrm{SC} = 11$ & $6\times 10^{-15}$ & 152.22 & 241\\
         STAR~\cite{akahoshi_partially_2023,sethi_rescq_2025} & \rz{} & $d_\mathrm{SC} = 7$ & $3.2\times 10^{-8}$ & 16.45 & 194\\
         \bottomrule
    \end{tabular*}
    \caption{Code distances, logical error rates and expected number of time-steps for different factory types considered in this work (at $p = 10^{-4}$). $d_X, d_Z, d_M$ are the distances of the surface code for the X basis, Z basis and measurement, respectively. CC stands for colour code, and SC stands for surface code. $\#q$ is the number of physical qubits needed for each factory, including the qubits needed for syndrome extraction.}\label{table:factory-error-rates}
\end{table}

\begin{figure}[htbp!]
    \centering
    \includegraphics[width=\linewidth]{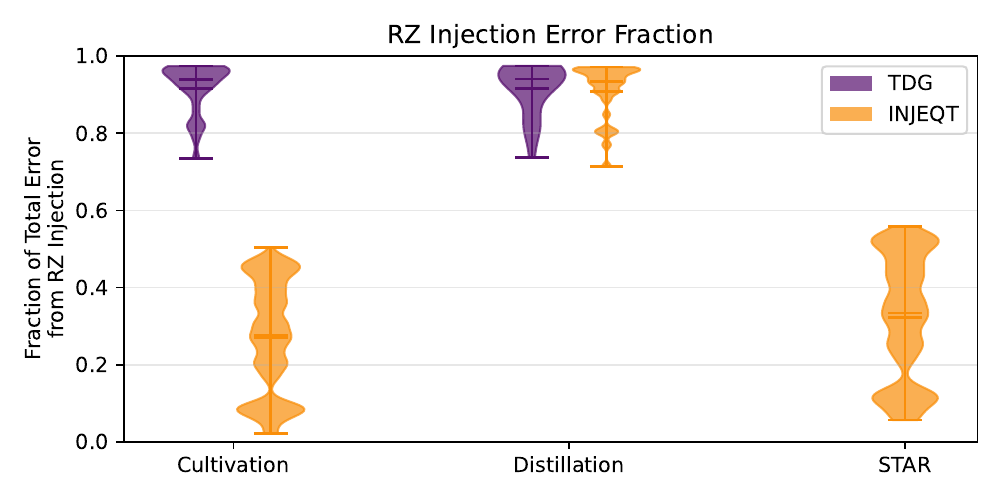}
    \caption{Fraction of the total program infidelity (error) that comes from \rz{} injections across all benchmarks. \acronym{} reduces the error contribution from \rz{} for all factory types.}\label{fig:rz-fraction}
\end{figure}

\par The discard rate for distillation is $\sim 5.5\times 10^{-3}$, and therefore we model distillation as a deterministic state preparation procedure. In contrast, cultivation and STAR have significantly higher discard rates (up to $25\%$), and thus their state preparations are modelled stochastically. Unlike distillation and cultivation, STAR directly prepares \rz{} states which can then be injected into the modules. We show the fraction of the program error rate that comes from these \rz{} injections for different factories in Figure~\ref{fig:rz-fraction}. Since the \tdg{} compilation does not consider direct injections of \rz{} states, we do not have a corresponding data point for STAR. However, we see that for the naive compilation scheme used by \tdg{}, \rz{} injection dominates the error budget. However, using our proposed method \acronym{}, we significantly reduce this error contribution. Note that the reduction in error contribution is negligible for distillation, and both STAR and cultivation have a much smaller error contribution from \rz{} injections. We discuss the reason for the former in Section~\ref{sec:injeqt} and for the latter below.

\begin{figure}[htbp!]
    \centering
    \includegraphics[width=0.9\linewidth]{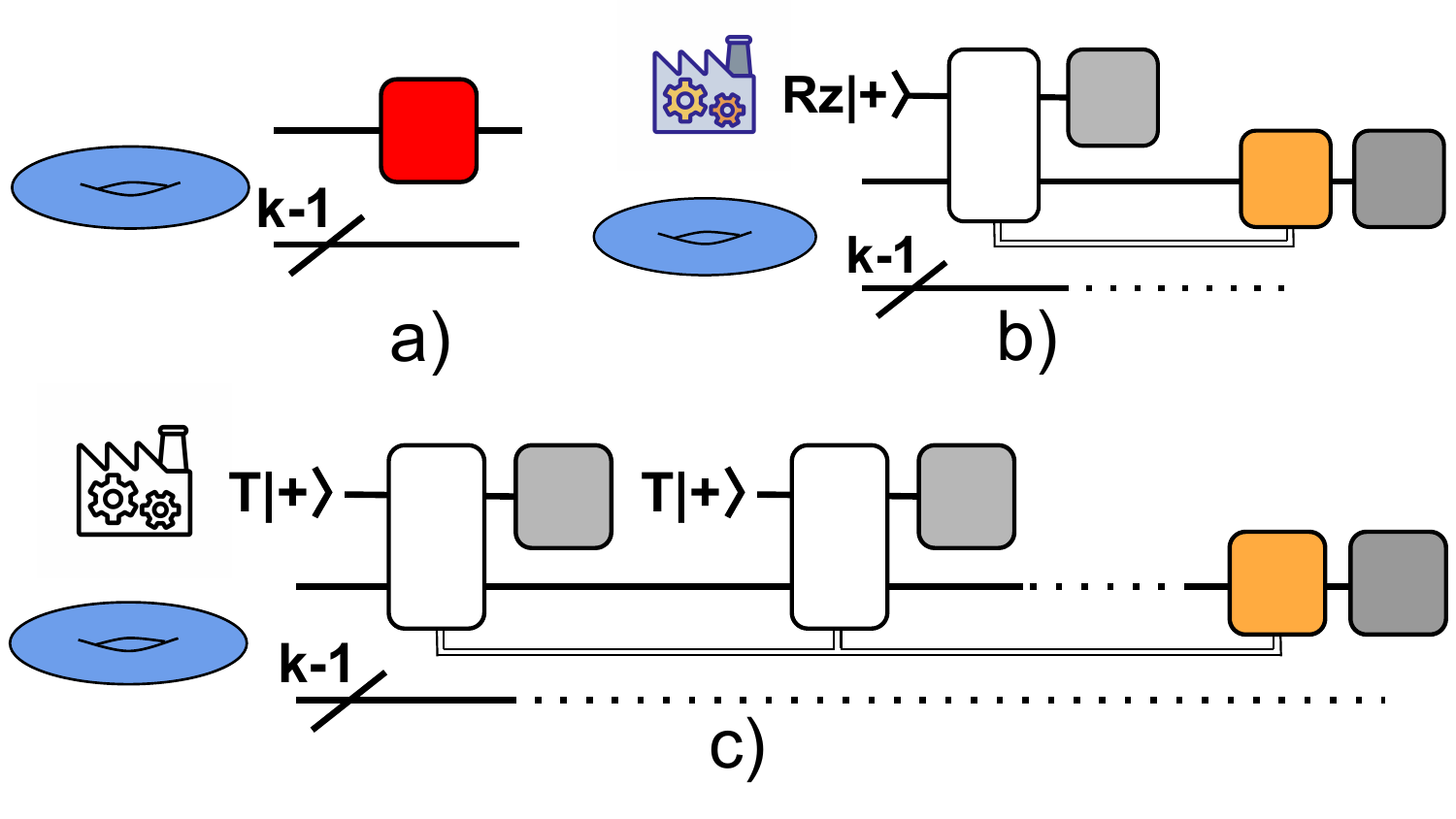}
    \caption{Illustration of magic state injection techniques within the Tour de Gross (TDG) ISA. (a) Magic state preparation represented as a primitive within the TDG instruction set architecture. (b) Direct injection of an Rz magic state sourced from a dedicated Rz-factory, followed by necessary feed-forward corrections. (c) Sequential injection of magic states as a series of T-gates from a T-factory to realize non-Clifford rotations, including subsequent adaptive Clifford correction.}
    \label{fig:injection-protocols}
\end{figure}

\par We compare the two injection protocols in Figure~\ref{fig:injection-protocols}. For every $\pauliphi$ gate that is approximated with precision $\varepsilon_\mathrm{synth}$ via GridSynth~\cite{ross_optimal_2016}, we require $c \approx -10\log(\varepsilon_\mathrm{synth})$ $\tstate$ injections. In \tdg{}, each of these injections requires a $\tstate$ factory preparation ($\tprep_{\tstate}$) and an inter-module measurement. Therefore, the total error incurred by this procedure is
\[\varepsilon_{\tstate} \approx c(\varepsilon_{\tprep_{\tstate}} + \varepsilon_{\intermodule}),\]
where we use the first order approximation. If we instead use a \rz{} factory ($\tprep_{\rz}$) for the $\pauliphi$ gate, we can directly inject the $\rz(\varphi)$ states, however, this requires non-Clifford correction of $\rz(2\varphi)$. If this correction fails, then we require $\rz(4\varphi)$ correction, and so on. Since each correction is required with $50\%$ probability, the expected number of injections is given by,
\[\mathbb{E}(\#\mathrm{injections}) = \sum_{i=1}^\infty i\cdot\frac{1}{2^i} = 2.\]
We can now compute the expected error up to the first order for the \rz{} injection version as
\begin{equation}
    \varepsilon_{\rz} \approx 2(\varepsilon_{\tprep_{\rz}} + \varepsilon_\intermodule).
    \label{eq:rz-error}
\end{equation}
Therefore, since we see $\varepsilon_{\rz} < \varepsilon_{\tstate}$ from the plots, we can derive the condition required to achieve this as,
\begin{equation}
    \begin{aligned}
        \varepsilon_{\rz} &< \varepsilon_{\tstate} \implies 2(\varepsilon_{\tprep_{\rz}} + \varepsilon_{\intermodule}) < c(\varepsilon_{\tprep_{\tstate}} + \varepsilon_\intermodule)\\
        &\implies \varepsilon_{\tprep_{\rz}} < \frac{c}{2}\left(\varepsilon_{\tprep_{\tstate}} + (1-2/c)\varepsilon_\intermodule\right)\\
        &\implies \varepsilon_{\tprep_{\rz}} < c\cdot\max\left(\varepsilon_{\tprep_{\tstate}}, \varepsilon_\intermodule\right).
    \end{aligned}
    \label{eq:rz-constraint}
\end{equation}
Note that for most applications, $\varepsilon_\mathrm{synth}$ will be of the order of $10^{-8}-10^{-10}$ (or even lower for deeper circuits), which gives us $80 \lessapprox c \lessapprox 100$. Therefore, as long as $\varepsilon_{\tprep_{\rz}}$ is comparable to, or even $1-2$ orders of magnitude worse in error rates than the larger of the inter-module or $\tstate$-factory error rates, we will have smaller error contributions from \rz{} factories. This condition holds for the STAR factory considered in this work. This motivates the case for direct \rz{} synthesis, despite it's non-determinism, an approach that has gained increasing attention in recent works~\cite{sethi_rescq_2025,yoshioka_transversal_2025,toshio_practical_2024,toshio_star-magic_2026}.
\section{Our Proposal: \acronym{}}\label{sec:proposal}

To improve program fidelity, or equivalently reduce program error rates, we can use alternative compilation techniques that trade more erroneous operations for less erroneous operations, at the cost of either space or time, or both. In this work, we propose, \acronym{}, wherein, we trade-off the most erroneous ISA instruction, i.e., the inter-module measurements, for more $\tstate$-factory instructions. We achieve this by adding a second-level factory that prepares \rz{} states using the $\tstate$-factory on an auxiliary code, and then inject the \rz{} state directly (Section~\ref{sec:injeqt}). In this work, we consider the surface code as the auxiliary code. However, this comes at the cost of increased execution times and space overhead. We then address this time overhead by pre-fetching the correction \rz{} states in parallel, which reduces the overall space-time cost of \acronym{}. This approach provides the best of both worlds in terms of improved error rates, and better space-time costs than \tdg{} (Section~\ref{sec:prefetch}).

\subsection{Improved Logical Error Rates for Injection}\label{sec:injeqt}

The motivation for building a second level factory for \rz{} injections follows from~\eqref{eq:rz-constraint}, where we observe that as long as we can construct \rz{} factories with comparable error rates, we expect an improvement in the total program error rate. We illustrate the working of our $2$-level factory in Figure~\ref{fig:injeqt-factory}. We can now compute the expected error rate of \acronym{}, $\varepsilon_\mathrm{\acronym}$, as,
\begin{equation}
        \varepsilon_\mathrm{\acronym} \approx 2\cdot (c\cdot (\varepsilon_{\tprep_{\tstate}} + \varepsilon_{\mathrm{tech}}) + \varepsilon_{\intermodule}),
\end{equation}
where $\varepsilon_\mathrm{tech}$ is the error rate of injection for lattice surgery, or transversal CNOT depending on the choice of technology used. Comparing this with~\eqref{eq:rz-error}, we get the \rz{} synthesis error rate, $\varepsilon_{\rz}$, of \acronym{} is $c\cdot (\varepsilon_{\tprep_{\tstate}} + \varepsilon_{\mathrm{tech}})$. We can further approximate this to the leading term, $c\cdot\varepsilon_{\tprep_{\tstate}}$ since the error rates for injections in both technologies are much smaller that the error rate of the $\tstate$-factories for reasons discussed in Section~\ref{sec:ms-injection}. We can now compare this to the inequality obtained in~\eqref{eq:rz-constraint} and we observe that this constraint is satisfied for the choice of factories considered in this work. However, note that for distillation, $\varepsilon_{\tprep_{\tstate}}$, is almost as large as $\varepsilon_\intermodule$, which gives us the limiting case of our inequality. This is why the improvement in the error fraction for distillation is much smaller than cultivation (Figure~\ref{fig:rz-fraction}). On the other hand, we notice a significant reduction in the error fraction for cultivation since $\varepsilon_{\tprep_{\tstate}}$ has a much lower error rate than $\varepsilon_\intermodule$ for cultivation.

\begin{figure}
    \centering
    \includegraphics[width=\linewidth]{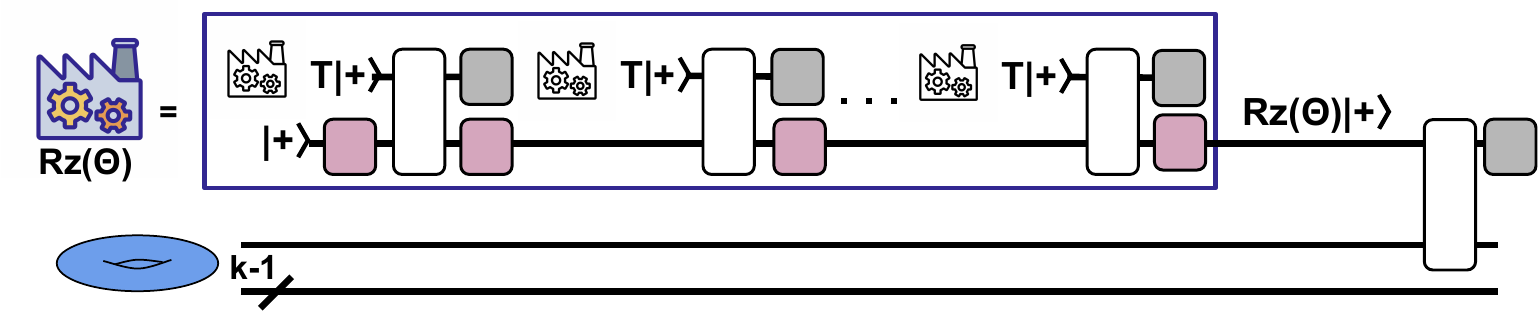}
    \caption{Execution of the $2$-level \acronym{} factory where we perform multiple sequential injections onto the auxiliary code. Each injection can be done either via lattice surgery or transversal CNOT (Figure~\ref{fig:ms-injections}). Once the $\rz(\theta)$ state is prepared in this auxiliary code, it is then injected onto the extractor module via an inter-module measurement.}
    \label{fig:injeqt-factory}
\end{figure}

\par We now compare the execution time for each $\pauliphi$ rotation between \acronym{} and \tdg{}. Since our proposal modifies only the injection part of the compilation, the execution time for in-module and inter-module measurements during the setup phase (Figure~\ref{fig:extractor-compilation}) remains unchanged. Therefore, we focus only on the differences in injection times. Each $\tstate$ injection in \tdg{} requires $c$ $\tstate$-factory preparations and $c$ inter-module measurements. Because \tdg{} uses a single factory, these operations are all sequential, except for the first $\tstate$ preparation which can happen simultaneously while the prior in-module and inter-module operations are being executed. This gives us $\tau_\mathrm{\tdg}$ as,
\[\tau_\mathrm{\tdg} = (c-1) \tau_{\tprep_{\tstate}} + c\tau_{\intermodule} \approx c(\tau_{\tprep_{\tstate}} + \tau_{\intermodule}).\]
We can also compute the expected execution time for \acronym{}, $\tau_\mathrm{\acronym}$ as,
\[\tau_\mathrm{\acronym} = 2\left(c(\tau_{\tprep_{\tstate}} + \tau_\mathrm{tech}) + \tau_{\intermodule}\right) = 2(\tau_\mathrm{prep} + \tau_\intermodule),\]
where $\tau_\mathrm{tech}$ is the time it takes for each injection from the $\tstate$-factory onto our second-level \acronym{} factory. $\tau_\mathrm{prep} = c(\tau_{\tprep_{\tstate}} + \tau_\mathrm{tech})$ is the preparation time for our second level \acronym{} factory. We now look at the ratio between these two approaches, $f_\mathrm{\acronym}$ as,
\begin{equation}
    f_\mathrm{\acronym} = \frac{\tau_\mathrm{\acronym}}{\tau_\mathrm{\tdg}} = 2\left(\frac{\tau_{\tprep_{\tstate}} + \tau_\mathrm{tech} + \tau_{\intermodule}/c}{\tau_{\tprep_{\tstate}}+\tau_{\intermodule}}\right) = 2\alpha,
    \label{eq:time-ratios}
\end{equation}
where $\alpha$ represents the variable part of the ratio. Assuming a sequential first preparation for \acronym{}, we observe that for the choice of technologies considered, $\tau_\mathrm{tech} < (1-1/c)\tau_\intermodule$, and therefore, we have $\alpha < 1$. However, since $\tau_{\tprep_{\tstate}} \gtrapprox \tau_\intermodule$ for both distillation and cultivation, we also have $\alpha \gtrapprox 0.5$, indicating that in expectation, we will always see an increase in execution time using a naive implementation of \acronym{}.

\begin{figure}[htbp!]
    \centering
    \includegraphics[width=\linewidth]{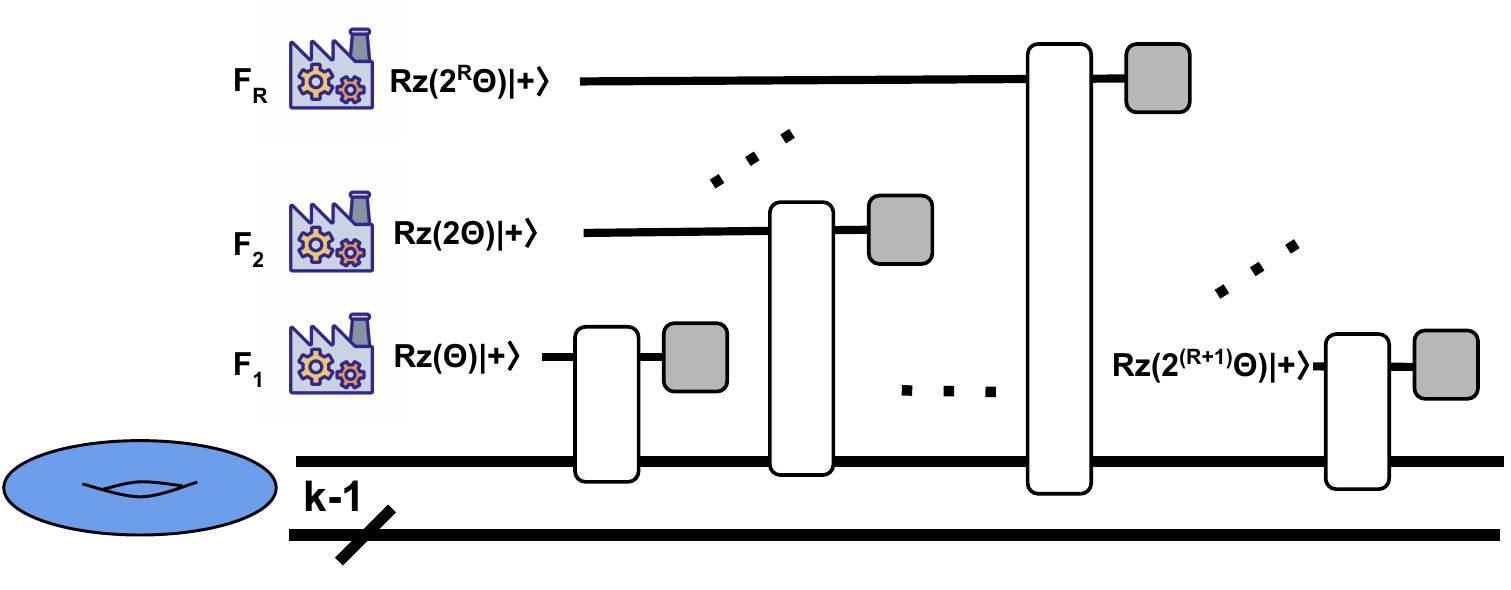}
    \caption{\acronym{} pre-fetching technique using $R$ copies of the $2$-level factories. All $R$ factories begin preparing different $\rz(\theta_i = 2^{i-1}\theta)$ states. Once the $\rz(\theta)$ state completes preparation in $F_1$, it is injected onto the extractor module. If the measurement result indicates that a correction is required, then the already prepared $\rz(2\theta)$ state is injected, and so on. Simultaneously, $F_1$ begins preparing the next un-prepared state, i.e., the $\rz(2^{R+1}\theta)$ state. The execution \textit{may} stall if the preparation time for $F_1$ exceeds the time taken to inject the previous $R$ states.}
    \label{fig:injeqt-prefetch}
\end{figure}

\subsection{Trading Time-Overheads for Space}\label{sec:prefetch}

\begin{figure*}
    \centering
    \includegraphics[width=\linewidth]{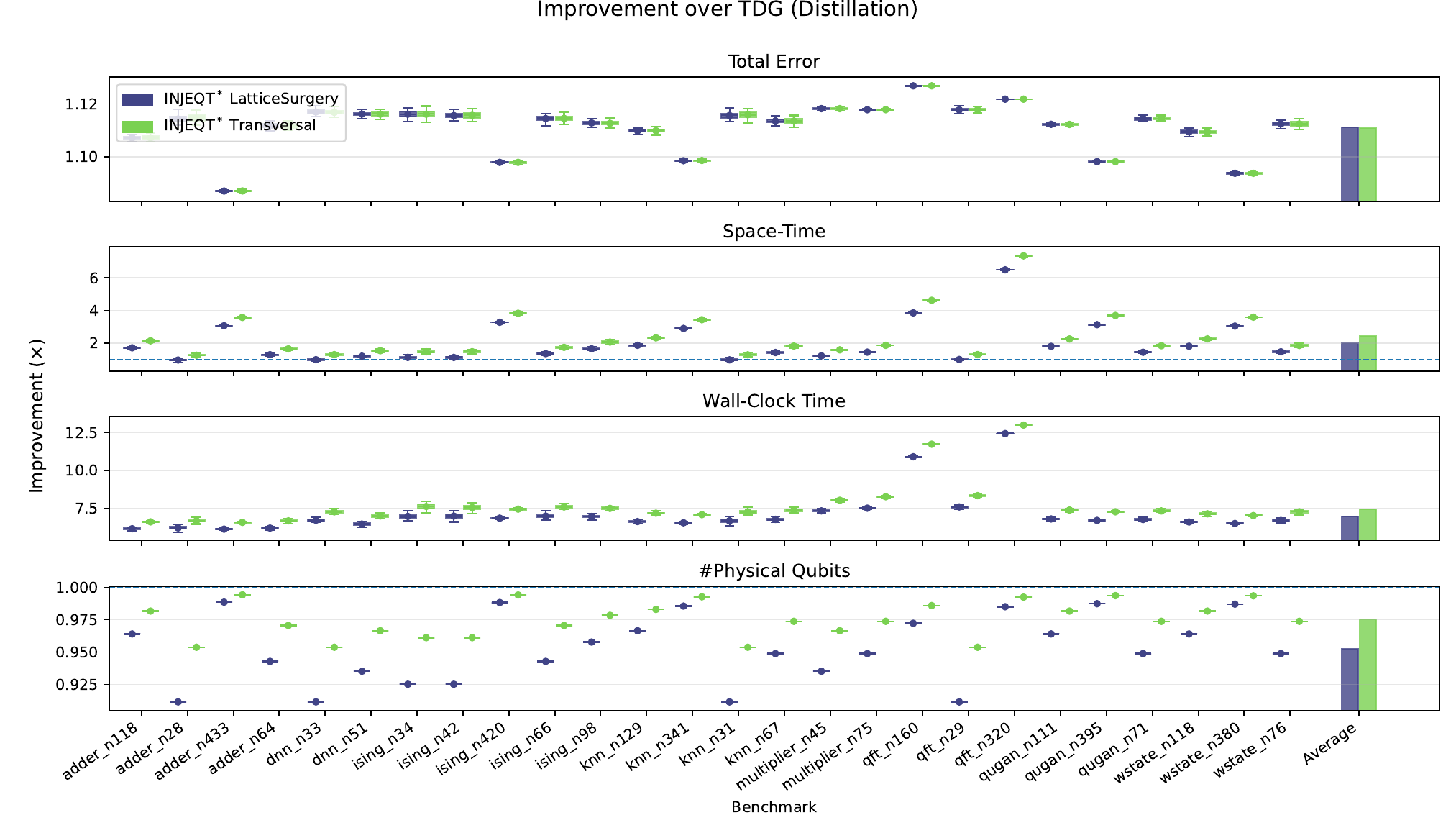}
    \caption{Improvements gained by \acronym{}$^*$ over \tdg{} across different metrics for the entire benchmark suite for the distillation factory. We also report the mean improvement at the end.}\label{fig:distillation-all}
\end{figure*}

Since $\rz(\varphi)$ injection fails with $50\%$ probability, we will need to inject the correction state, $\rz(2\varphi)$, with $50\%$ probability. Therefore, we can reduce the total execution time by \textit{pre-fetching}, i.e., preparing these correction states in advance, since each of the correction states will be required with some non-zero probability. We optimize the $2$-level \acronym{} factory by \textit{pre-fetching} multiple $\rz(\theta_i)$ states in parallel for $\theta_i = 2^{i-1}\varphi$, where $i \in [1, R]$, where we use $R$ factories. We illustrate this approach in Figure~\ref{fig:injeqt-prefetch}. Furthermore, once any of these factories are used for injection, we start preparing the next correction angle that is not being prepared by any of the other factories. We can now compute the expected execution time, $\tau_{\mathrm{\acronym}^\infty}$, assuming we have sufficiently many factories so that we do not experience stalls as,
\begin{equation}
    \begin{aligned}
        \tau_{\mathrm{\acronym}^\infty} = &c(\tau_{\tprep_{\tstate}} + \tau_\mathrm{tech}) + 2\tau_\intermodule = \tau_\mathrm{prep} + 2\tau_\intermodule.
    \end{aligned}
    \label{eq:injeqt-opt}
\end{equation}
If we do not have sufficiently many factories, i.e., $R < 1 + \tau_\mathrm{prep}/\tau_\intermodule$, despite prefetching, we will need to wait for the preparation to complete every $R$ injections. However, since injections beyond the first two injections become quite unlikely, we can ignore the effect of stalls. This is indeed what we see in Section~\ref{sec:sweep}, where the improvements in wall clock time (total execution time of the program) quickly saturates for a small number of factories.

\par We can further optimize this pre-fetching time by starting the pre-fetching in parallel with the \textit{setup} phase of the extractor modules, i.e., when the extractor modules are executing in-module and inter-module measurements. Since the setup phase requires $\sim 18.5$ sequential in-module on average and about $O(\log M)$ inter-module operations to prepare the GHZ state, where we have $M$ modules on device. Therefore, even though $\tau_\mathrm{prep}$ is large due to $c$ being large, a large part of it gets hidden due to the large time it takes during the extractor setup phase. This further strengthens our pre-fetching approach since it reduce the time overhead of preparing all states in advance. Therefore, in the optimistic case when we can pre-fetch in advance, we will have $\tau_{\mathrm{\acronym}^\#} \to 2\tau_\intermodule$, where $\tau_{\mathrm{\acronym}^\#}$ is the theoretical optimal execution time of \acronym{}.

\begin{figure*}
    \centering
    \includegraphics[width=\linewidth]{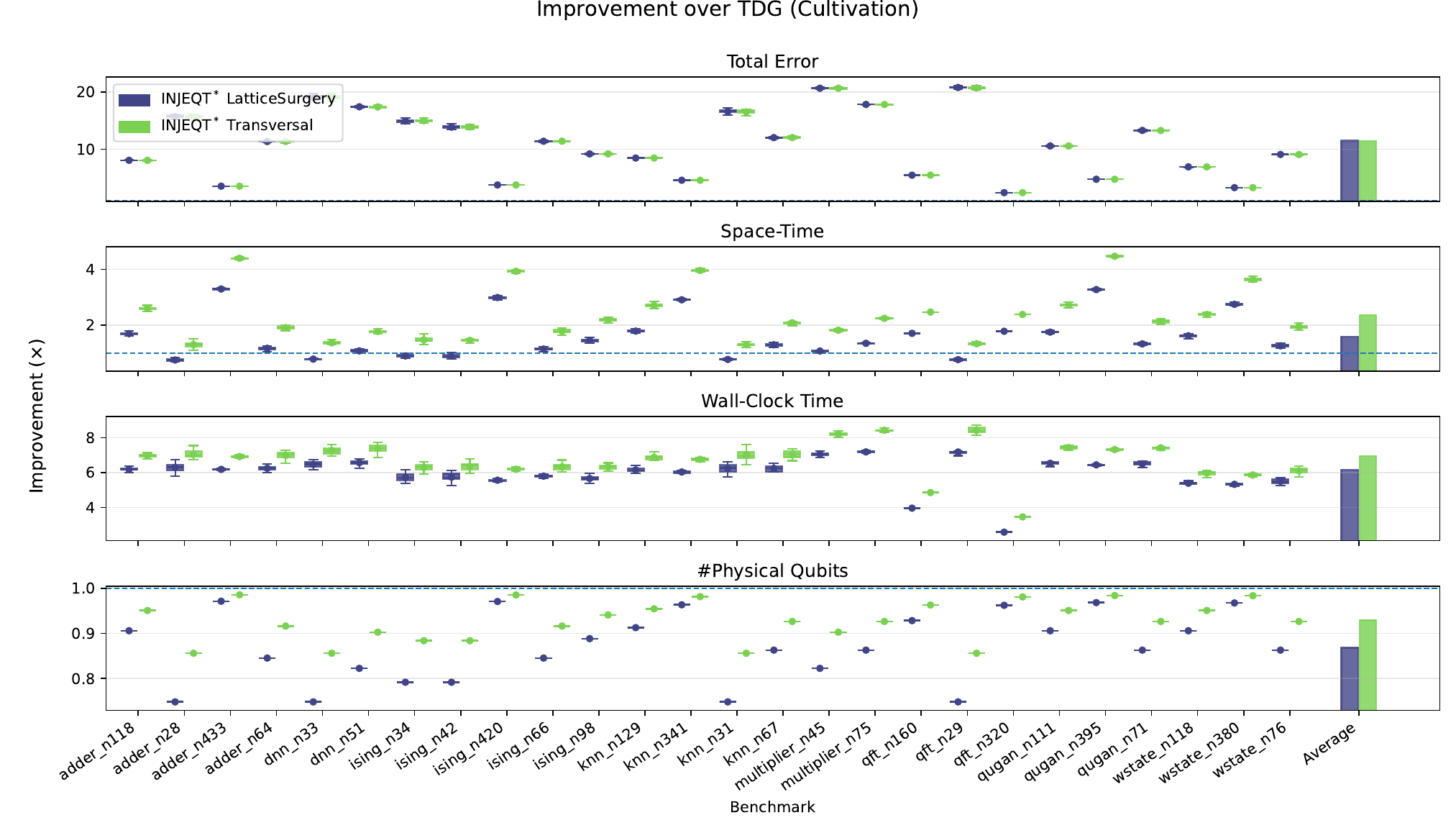}
    \caption{Improvements gained by \acronym{}$^*$ over \tdg{} across different metrics for the entire benchmark suite for the cultivation factory. We also report the mean improvement at the end.}\label{fig:cultivation-all}
\end{figure*}
\begin{figure*}
    \centering
    \includegraphics[width=\linewidth]{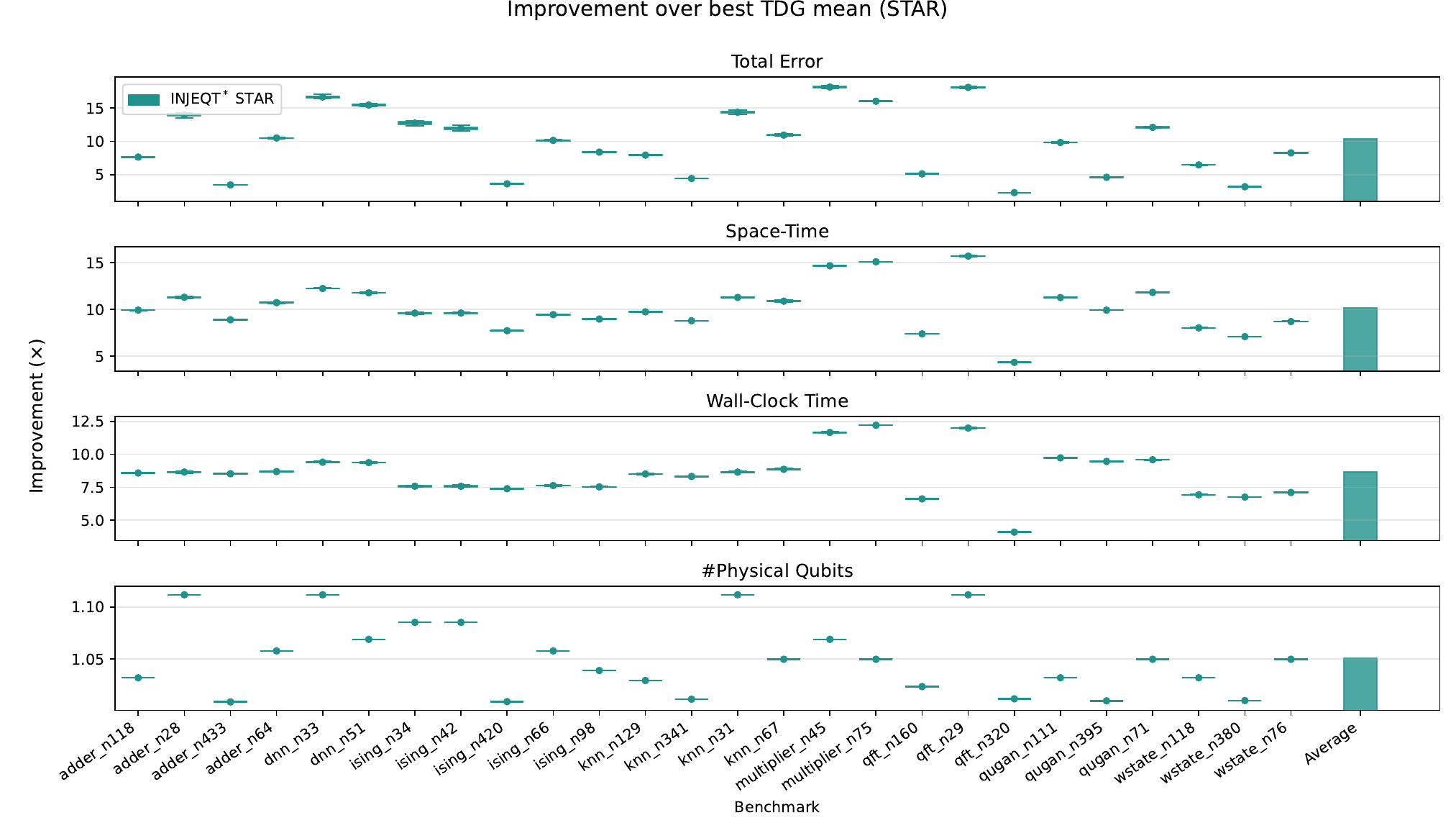}
    \caption{Improvements gained by \acronym{}$^*$ over \tdg{} across different metrics for the entire benchmark suite for the STAR factory. We use the better of distillation or cultivation for the corresponding \tdg{} metric. We also report the mean improvement at the end.}\label{fig:star-all}
\end{figure*}

\section{Evaluation}\label{sec:evaluation}

\subsection{Simulation Methodology}
We evaluate \acronym{} against the baseline \tdg{} compilation for the three kinds of factories considered in this work (Table~\ref{table:factory-error-rates}). For each input circuit from the QASMBench~\cite{li_qasmbench_2022} suite, we first compile it into extractor instructions as described in Section~\ref{sec:extractor-isa} and illustrated Figure~\ref{fig:extractor-compilation}. For \tdg{}, we compile magic-state injections into $\tstate$ injections via GridSynth~\cite{ross_optimal_2016}. For \acronym{}, we instead compile them into $\rz(\theta)$ injections and model the injection failures using seeded execution and run $20$ trials, computing metrics for each trial. For cultivation and STAR, we also model factory preparation discards, and sample factory preparation times by modelling the repeat-until-success attempts during execution.
\par For both policies, we compute the total program error rate using a first order approximation by adding up the error contributions of individual operations. We neglect accounting for the trivial operations whose \textit{cumulative} error contribution is not large enough to be noticeable up to the fifth decimal place (syndrome extraction, automorphisms). The relative frequencies of these operations do not offset the relative disparities in error rates of inter and in module measurements. For each trial, we compute the wall-clock time of program execution, which is total number of cycles required to execute the program. We also report the number of physical qubits needed for each benchmark, accounting for factory qubits (including additional qubits used by \acronym{} for the $2$-level factory), extractor modules, adapters and LPUs (logical processing units). Finally, we compute the space-time cost of each benchmark, which is the product of the wall-clock time and the number of physical qubits. For each of these metrics, we plot the ratio of the metric for \acronym{} over \tdg{}, and report that as `$\times$ improvement'.
\par We evaluate our $2$-level factory in \acronym{} for both lattice surgery and transversal injection models. Additionally, for \acronym{}, we compute these metrics for $R \in [1, 20]$ and pick the value of $R$ that optimizes the particular choice of metric and report the choice as \acronym{}$^*$ for the plots in Section~\ref{sec:eval-factories}. We discuss how this choice of factories affects these metrics further in Section~\ref{sec:sweep}.

\begin{figure*}
    \centering
    \begin{subfigure}{0.49\linewidth}
        \centering
        \includegraphics[width=\linewidth]{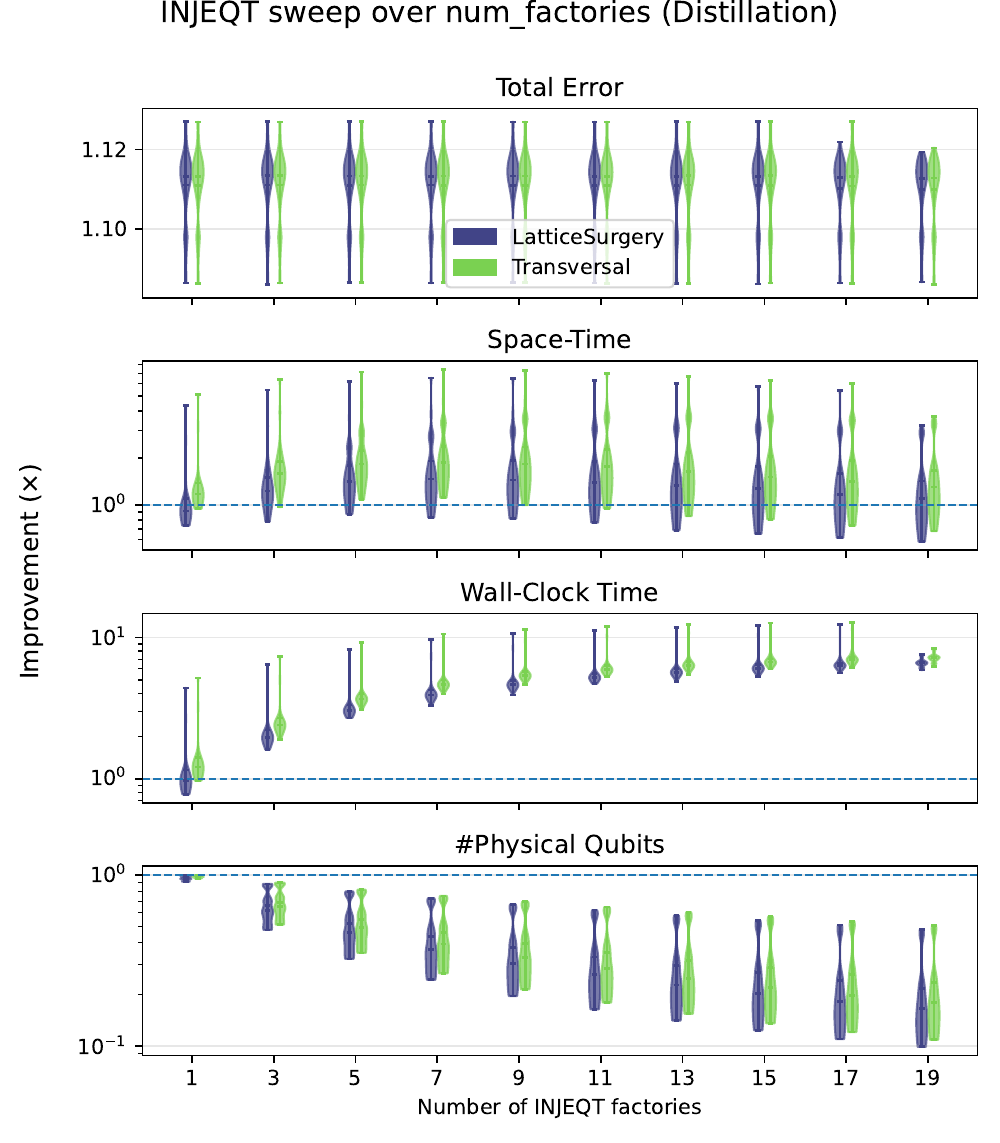}
        \caption{Distillation}\label{fig:distillation-sweep}
    \end{subfigure}
    \begin{subfigure}{0.49\linewidth}
        \centering
        \includegraphics[width=\linewidth]{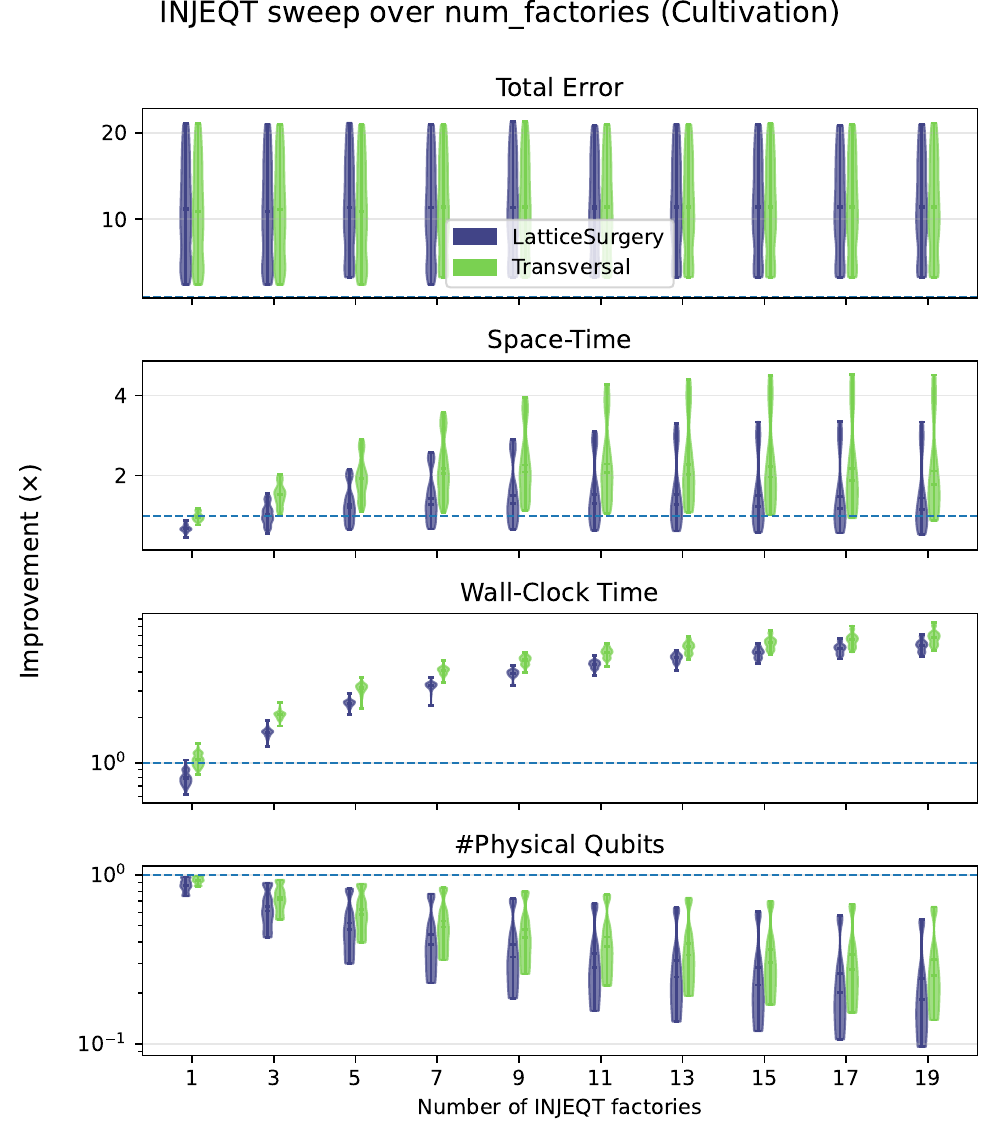}
        \caption{Cultivation}\label{fig:cultivation-sweep}
    \end{subfigure}
    \caption{Sweep over different choices of \acronym{} factories as a violin plot across all benchmarks and trials as an improvement over \tdg{} compilation for distillation and cultivation.}
\end{figure*}

\subsection{Evaluation against Different Factories}\label{sec:eval-factories}

\subsubsection{Distillation}

We plot the improvement obtained by \acronym{}$^*$ for different metrics in Figure~\ref{fig:distillation-all}. We observe up to a $12.5\%$ improvement in error rates, with an average improvement of over $11\%$. Note that the error rate $\varepsilon_{\tprep_{\tstate}}$ for distillation is only slightly lower than the inter-module measurements, making this factory the limiting case for \acronym{}. As factory error rates get better, improvements gained by employing \acronym{} will get better. We also observe up to a $13\times$ improvement in the wall-clock time with an average improvement of about $7\times$. As we will discuss in Section~\ref{sec:sweep}, getting the optimal improvement does not necessarily require $R$, the number of \acronym{} factories, to obtain these best improvements. The increase in the number of physical qubits is only at most $9\%$ with a single $2$-level factory with an average increase of $5\%$ for the lattice surgery approach and $2.5\%$ increase for the transversal approach. In space-time, we see up to a $7.2\times$ improvement with an average improvement of about $2\times$.

\subsubsection{Cultivation}

We plot the improvement obtained by \acronym{}$^*$ for different metrics in Figure~\ref{fig:cultivation-all}. We observe up to a $20\times$ improvement error rate with an average improvement of $12.5\times$. We observe up to a $8.8\times$ improvement in the wall-clock time with an average improvement of $6.5-7\times$. Using a single \acronym{} factory on top of cultivation leads to up to a $25\%$ increase in the number of physical qubits with an average increase of $13\%$ for lattice surgery and a $7\%$ increase for transversal architectures. On space-time, we see up to a $4.5\times$ improvement and an average improvement of about $1.5\times$ for lattice surgery and $2.5\times$ for transversal.
\par Unlike distillation where its error rate was only marginally better, the choice of cultivation factory in this work has a much lower error rate, $\varepsilon_{\tprep_{\tstate}}$, than the intermodule measurements, this also indicates the base-case improvement in error obtainable by \acronym{}. On the other hand, we see a larger increase in the space overhead since cultivation uses fewer qubits, making the space overhead of the $2$-level \acronym{} factories more significant. We also see smaller improvements in time since cultivation has larger factory preparation times, which obscures the gains from pre-fetching and parallel preparations of the $\rz(\theta_i)$ states in \acronym{}.

\subsubsection{STAR}

Since \tdg{} does not consider \rz{} factories, when comparing \acronym{} that uses STAR factories, we instead choose the choice of \tdg{} compilation between distillation and cultivation that performs better on the metric being compared against. We report these results in Figure~\ref{fig:star-all}. Unlike distillation and cultivation, we do not require a second-level factory for STAR since it directly prepares \rz{} states. In addition to this, since the cycle time for the direct preparation is small, additional \acronym{} factories do not lead to any meaningful gains in the execution time, making a single factory suffice. We observe up to an $18\times$ improvement and an average improvement of about $10\times$ in error rates. We improve by up to $12\times$ and about $9\times$ on average on the wall-clock time. Since STAR has a smaller physical qubit footprint, we also see up to a $10\%$ improvement and about a $4\%$ average improvement in the number of physical qubits. In space-time, we see up to a $16\times$ improvement with about a $12\times$ average improvement.

\begin{figure}
    \centering
    \includegraphics[width=\linewidth]{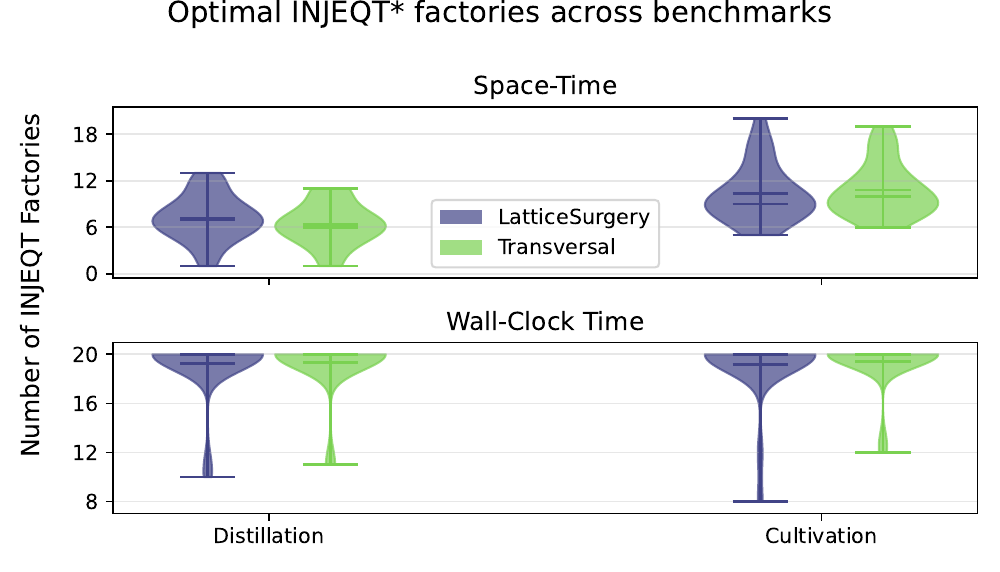}
    \caption{Violin plot showing the optimal choice of number of \acronym{} factories that optimizes the mean space-time cost and the wall-clock time for each benchmark.}\label{fig:num-factories-all}
\end{figure}

\subsection{Sensitivity to Number of \acronym{} Factories}\label{sec:sweep}
Since the evaluation in Section~\ref{sec:eval-factories} reported the improvement using the best choice of factories, we also perform a sensitivity sweep over the number of factories and report the results in Figure~\ref{fig:distillation-sweep} for distillation and Figure~\ref{fig:cultivation-sweep} for cultivation. We see no change in the total error irrespective of the number of factories since increasing the number of factories does not affect the total number of meaningful operations being executed by the quantum system. Additional factory preparations are discarded if unused, and do not affect the total error rate of the program. Additionally, as mentioned earlier, each additional factory adds more physical qubits and therefore we see an increase in the number of physical qubits, with up to a $10\times$ increase for $20$ \acronym{} factories. However, as we start to run larger quantum algorithms with more logical qubits, the fraction of space occupied by factories will reduce, thereby reducing this increase in space from \acronym{}. 
\par As discussed in Section~\ref{sec:prefetch}, each additional factory leads to improvement in the wall-clock time, however this begins to saturate quickly with a small number of \acronym{} factories, reducing the space footprint needed to observe maximal improvement in wall-clock time. Note that the result derived in~\eqref{eq:time-ratios} uses the expected execution time and the actual execution time varies. Therefore, runs of the quantum application when the injections succeed in the first try and do not require corrections, will have $f_\mathrm{\acronym} < 1$. Conversely, for runs that experience more injection errors will take increased time per $\pauliphi$ gate, leading to $f_\mathrm{\acronym} > 2$. As a result, even with $1$ \acronym{} factory, we see comparable wall-clock times which are sometimes better and sometimes worse than \tdg{}. For space-time, the results become more interesting where we see that there is no single good choice for the number of \acronym{} factories ($R$) across all benchmarks and there is always a small subset of benchmarks that performs worse than \tdg{} on space-time for a fixed choice of $R$.
\par To better understand this, we also report the number of \acronym{} factories needed to observe the improvements in wall-clock time and space-time reported as \acronym{}$^*$ in Figure~\ref{fig:num-factories-all}. We observe that although most benchmarks require $R$ close to $20$ to obtain the optimal improvement in wall-clock time, for a small fraction of benchmarks, even $R = 10$ suffices. Additionally, for space-time cost, we see that there is a wide ranging distribution for the optimal choice of $R$, with the requirement being lower for distillation since distillation requires lesser time for $\tstate$ preparations than cultivation, which lowers the requirement for $R$ needed that avoids stalling the preparations despite pre-fetching.
\section{Conclusion}
Extractor-based architectures are a promising direction on the path towards Fault-Tolerant Quantum Computing (FTQC) due to their high spatial-efficiency. However, program error rates in these architectures are bottlenecked by the most expensive instruction, the inter-module measurements. Performing universal quantum computation on these architectures require injecting resource-states via expensive inter-module measurements which contributes to over $90\%$ of the total program error. We propose \acronym{}, a novel $2$-factory design for resource-state injection. \arxiv{\footnote{We would like to note that recently we became aware of Ref.~\cite{liu_assessing_2026}, which studies a related problem under their Opt-1 optimization. Here, we propose an alternative pipelining strategy that further improves wall-clock time compared to their technique. Additionally, we evaluate the performance of \acronym{} across multiple factory paradigms (distillation, cultivation, and STAR), as well as for different auxiliary-code execution models.}} Our approach first prepares \rz{} states in an auxiliary code (e.g., the surface code), and then injects the prepared state into the extractor modules. Although this injection protocol requires another \rz{} correction (with $50\%$ probability), \acronym{} improves program error rates by up to $12.5\%$ for distillation factories, up to $22\times$ for cultivation factories, and up to $18\times$ for STAR factories. 

To address the additional latency introduced by auxillary synthesis, we further propose a pre-fetching technique that prepares the \rz{} state and its potential corrections states (that may be needed) in parallel, with ongoing in-module and inter-module operations. This approach improves the wall-clock times by up to $13\times$ for distillation, $8.8\times$ for cultivation and $12\times$ for STAR. Moreover, we reduce the space-time cost by up to $7.2\times$ for distillation, $4.5\times$ for cultivation, and $16\times$ for STAR. Finally, we evaluate \acronym{} for both lattice surgery and transversal injections demonstrating that \acronym{} is robust across factory choices and qubit technologies, paving the way for more efficient FTQC system designs.

\arxiv{
\section*{Acknowledgements}
This work was funded in part by the Texas Quantum Institute (TQI), and the NSF 24-599: Quantum Leap Challenge Institutes (QLCI), and the NSF STAQ project (PHY-2325080). Support is also acknowledged from the U.S. Department of Energy, Office of Science, National Quantum Information Science Research Centers, Quantum Systems Accelerator (Award No. DE-SCL0000121).
}

\bibliographystyle{IEEEtranS}
\bibliography{references}

\end{document}